\documentclass[iop,apj,tighten,a4paper]{emulateapj}

\usepackage{amsmath,amstext,amssymb}
\usepackage[breaklinks,colorlinks,citecolor=blue,linkcolor=magenta]{hyperref}

\def\kms{{\rm km\,s^{-1}}}

\def\msun{{\rm M_{\odot}}}
\def\msunkpc2{{\rm M_{\odot}\,kpc^{-2}}}

\slugcomment{Accepted for publication in ApJ}
\shorttitle{CDM Substructures in Early-Type Galaxies}
\shortauthors{Fiacconi et al.}

\begin{document}

\title{Cold Dark Matter Substructures in Early-Type Galaxy Halos}

\author{Davide Fiacconi\altaffilmark{1}, Piero Madau\altaffilmark{1,2,3}, Doug Potter\altaffilmark{1}, and Joachim Stadel\altaffilmark{1}}
\email{fiacconi@physik.uzh.ch}
\altaffiltext{1}{Center for Theoretical Astrophysics and Cosmology, Institute for Computational Science, University of Zurich, Winterthurerstrasse 190, CH-8057 Z\"{u}rich, Switzerland}
\altaffiltext{2}{Institute for Astronomy, ETH Zurich, CH-8093 Z\"{u}rich, Switzerland}
\altaffiltext{3}{Department of Astronomy \& Astrophysics, University of California, 1156 High Street, Santa Cruz, CA 95064, USA}


\begin{abstract}
We present initial results from the ``Ponos'' zoom-in numerical simulations of dark matter substructures in massive ellipticals.
Two very highly resolved dark matter halos with $M_{\rm vir}=1.2\times 10^{13}\,\msun$ and $M_{\rm 
vir}=6.5\times 10^{12}\,\msun$ and different (``violent" vs. ``quiescent") assembly histories have been simulated down to $z=0$ 
in a $\Lambda$CDM cosmology with a total of 921,651,914 and 408,377,544 particles, respectively. Within the virial radius, the 
total mass fraction in self-bound $M_{\rm sub}>10^6\,\msun$ subhalos at the present epoch is 15\% for the violent 
host and 16.5\% for the quiescent one. At $z=0.7$, these fractions increase to 19 and 33\%, respectively, as more recently 
accreted satellites are less prone to tidal destruction. In projection, the average fraction of surface mass density in substructure at 
a distance of $R/R_{\rm vir}=0.02$ ($\sim$ 5--10 kpc) from the two halo centers ranges from 0.6\% to $\gtrsim 2$\%, significantly higher than 
measured in simulations of Milky Way-sized halos. The contribution of subhalos with $M_{\rm sub} < 10^9\,\msun$ to the projected 
mass fraction is between one fifth and one third of the total, with the smallest share found in the quiescent host. We assess the impact of 
baryonic effects via twin, lower-resolution hydrodynamical simulations that include metallicity-dependent gas cooling, star formation, 
and a delayed-radiative-cooling scheme for supernova feedback. Baryonic contraction produces a super-isothermal total 
density profile and increases the number of massive subhalos in the inner regions of the main host. The host density profiles and projected subhalo mass 
fractions appear to be broadly consistent with observations of gravitational lenses.
\end{abstract}


\section{Introduction}

Precision measurements across cosmic time have made the cold dark matter plus cosmological constant ($\Lambda$CDM) paradigm one of the pillars of our 
current understanding of the origin and evolution of structures in the universe. This model has proven 
to be remarkably successful at matching observations from scales that span the horizon length \citep[e.g.,][]{planck+15} all the way down to the scales 
probed by the Lyman-$\alpha$ forest \citep[e.g.,][]{viel+13}. 
A vast zoo of non-baryonic dark matter candidates has been proposed over the last 
three decades to reproduce the wealth of cosmological/astrophysical data \citep[e.g.,][]{feng10}. 

The standard theory of structure formation requires dark matter to be cold, i.e. made of particles that become non-relativistic well before the matter domination 
era, and therefore clump on all scales. It is precisely on the smallest subgalactic scales
that there have been persistent observational challenges to the cold, collisionless dark matter expectations. These ``small-scale controversies", predominantly 
found in the abundances and density profiles of dark matter-dominated dwarfs in the local universe, may simply stem from a poor understanding of the baryonic 
processes involved in galaxy formation \citep[e.g.][]{pontzen+14,madau+14,weinberg+15}. They may alternatively indicate the need for more complex physics in the 
dark sector, and many modifications of the properties of the dark matter particle have been proposed to suppress small-scale power and alleviate some of these 
problems, including warm dark matter \citep[WDM;][]{bode+01}, self-interacting dark matter \citep{spergel+00}, fuzzy dark matter \citep{hu+00}, 
and superWIMPS \citep{cembranos+05}.
 
There is, however, no consensus on how serious these problems really are for $\Lambda$CDM, and detailed testing of the standard paradigm 
on small scales remains one of the most pressing issues in cosmology. Numerical simulations in $\Lambda$CDM have shown a rich spectrum of substructures in 
galaxy halos, the fossil remnants of a hierarchical merging process that is never complete \citep{moore+99,klypin+99,diemand+07,diemand+08,springel+08,stadel+09}. 
\citet{metcalf+01} first showed that self-bound subhalos would have a readily detectable gravitational lensing effect if they were as numerous as predicted in $\Lambda$CDM.
In particular, small fluctuations in the galaxy-scale lensing potential caused by substructures should result in measurable perturbations in the relative 
magnifications of quadruply-lensed quasar images. Discrepancies between the observed flux ratios and those predicted by a smooth lens model (``flux ratio anomalies")
have indeed been found to be common in quasar lenses \citep[e.g.,][]{mao+98,chiba02,metcalf+02}, and can be explained if a large fraction ($\sim 2$\%) of the projected 
mass at the Einstein radius is in the form of local substructures \citep{dalal+02}. 

To date, there is still no general agreement on whether semi-analytic models or $N$-body simulations in $\Lambda$CDM predict enough substructures to 
explain the frequency of lens anomalies in currently available samples \citep[e.g.,][]{bradac+04,maccio+06,amara+06,xu+09,chen+11,metcalf+12,xu+15}.
The discrepancy largely arises because of the small number of subhalos anticipated to survive near the radii where images form (typically around 5--10 kpc in projection). 
Somewhat surprisingly, most numerical simulation work has focused on present-day Milky Way-sized halos as the lens system. And yet, most gravitational lens 
galaxies are known to be early-type massive galaxies in relatively low density environments \citep[e.g.][]{kochanek+00}, while the 
average deflector redshift of the multiply-imaged quasars in the Cosmic Lens All-Sky Survey (CLASS) is $\langle z_{\rm lens}\rangle\approx 
0.6$ \citep{browne+03}. The abundance of substructures in a host is set by the competition between tidal disruption and
new accretion. More massive hosts (as well as hosts at higher redshifts) are then expected to be more clumpy because their subhalos have
been accreted more recently and managed to survive tidal destruction \citep{zentner+05,klypin+11}.

In order to refine $\Lambda$CDM predictions for substructure lensing, we have initiated a program, dubbed ``Ponos", of 
very high resolution $N$-body 
and hydrodynamic cosmological simulations of early-type massive galaxies. In this Paper we present initial results on halo substructures from new 
``zoom-in" simulations of two $\sim 10^{13}\,\msun$ systems, each having between 155 and 286 million dark matter particles within the virial radius
$R_{\rm vir}$ (defined here as the radius enclosing a mean matter density $\Delta\,\rho_c$, where $\rho_c$ is the critical density and $\Delta$ 
is the redshift-dependent virial overdensity, \citealt{bryan+98}).
On these halo mass scales, the global fraction of E/S0 galaxies is estimated to be about 50\% \citep{wilman+12}.
We focus on very high resolution calculations of a limited number of halos to study in detail the relative impact of light and 
heavy subhalos in strong lensing produced by elliptical galaxies, while we plan to extend this study to a larger halo sample.
The two target halos are part of the AGORA Simulation Project\footnote{\url{sites.google.com/site/santacruzcomparisonproject/}.} \citep{kim+14}.


\section{Numerical simulations}

The Ponos collisionless simulations have been performed with the Tree/$N$-body code {\sc Pkdgrav3}. Gravity was computed using a fast multipole 
method \citep{dehnen+00,dehnen+02} based on a $5^{\rm th}$-order reduced expansion for faster and more accurate force calculation, 
and a multipole based Ewald summation technique for periodic boundary conditions \citep{stadel+09,stadel+13}. An opening 
angle of $\theta=0.55$ was adopted for the gravity tree. Time-steps were assigned individually to each particle based on the local 
dynamical time $\Delta t \leq \xi / \sqrt{G \rho}$, where $\xi = 0.03$ is an accuracy parameter and $\rho$ is the density enclosed 
within the particle orbit \citep{zemp+07}. This time-stepping strategy is both faster and 
more adaptive and accurate than the conventional time-step $\Delta t \leq \xi \sqrt{\epsilon_g/|\mathbf{a}|}$, 
where $|\mathbf{a}|$ is the magnitude of the local acceleration and $\epsilon_g$ is the gravitational softening. 

Initial conditions (ICs) were generated with the {\sc Music}\footnote{\url{www.phys.ethz.ch/~hahn/MUSIC.}} package \citep{hahn+11}, and are based upon a 
{\it Wilkinson Microwave Anisotropy Probe} cosmology with $\Omega_M=0.272$, $\Omega_\Lambda=0.728$, $\Omega_b=0.0455$, $\sigma_8=0.807$, $n_s=0.961$, 
and $H_0=70.2\,$km s$^{-1}\,$Mpc$^{-1}$ \citep{hinshaw+13}. The two target halos were selected from a low resolution pathfinder simulation of a 85.5 comoving Mpc 
box, and are characterized by very different assembly histories \citep{kim+14}. 
One (``PonosQ", $M_{\rm vir}=6.5\times 10^{12}\,\msun$)    
has a relatively {\it quiescent} early merger history (i.e. few major mergers between $z=4$ and 2, as defined by \citealt{kim+14}), while the other (``PonosV", $M_{\rm vir}=1.2\times 10^{13}\,\msun$)
is characterized by a more {\it violent} early merger history (i.e. many mergers between $z=4$ and 2) \footnote{We note that, despite the naming of the simulations, PonosQ has a more active assembly history at low redshift than PonosV, as discussed in Section \ref{halo_detection}.}.
A strong isolation criterion was imposed on the quiescent halo, one in which its $3 R_{\rm vir}$ radius circle does not intersect the $3 R_{\rm vir}$ radius circle of any 
halo with half or more of its mass at $z=0$. A 
relaxed criterion was used for the most massive violent halo: $2 R_{\rm vir}$ circle instead of $3 R_{\rm vir}$.    
Higher-resolution simulations were performed on new ICs re-centered on each of the target halos, for a total of six further nested spatial refinements (each by a factor of 
2), corresponding to an effective resolution of 8192$^3$ particles with mass $m_p=4.2\times 10^4\,\msun$. The final highest-resolution region is the convex hull Lagrangian
volume that contains all the particles that fall within $3R_{\rm vir}$ of the target halo at $z=0$, sufficiently large to include all the structures that merge with
the main host or have a significant impact on its evolution. The high resolution regions of PonosV and PonosQ were sampled with 890 and 388 million particles, respectively, and 
evolved with a force resolution $\epsilon_g=210$ pc (fixed in physical units after $z=9$ and equal to 1/50 of the initial interparticle separation) 
from $z=100$ to $z=0$. Contamination within $R_{\rm vir}$ from coarse-level particles 
was found to be well below 0.1\% in both number of particles and mass fraction down to the present epoch. To check for numerical convergence, both halos were re-simulated 
at a factor two lower spatial resolution. Some basic properties of the simulated target halos are listed in Table \ref{tab:summary}.

\begin{deluxetable*}{l l c c c c c c c c}
\tablecaption{Properties of the simulated elliptical halos.\label{tab:summary}}
\tablehead{
\colhead{Name} & \colhead{Merger History} & \colhead{$z_f$} & \colhead{$M_{\rm vir}$} & \colhead{$R_{\rm vir}$} & \colhead{$V_{\rm host}$} & \colhead{$r_s$} & \colhead{$c$} & \colhead{$N_{\rm vir}$} & \colhead{$N_{\rm sub}$}\\
\colhead{} & \colhead{($2<z<4$)} & \colhead{} & \colhead{$(\msun)$} & \colhead{(kpc)} & \colhead{$(\kms$)} & \colhead{(kpc)} & \colhead{} & \colhead{} & \colhead{}
}
\startdata
PonosV  & violent & 1.10 & $1.2\times 10^{13}$ & 600.5 & 348.2 & 50.9 & 11.8 & 285,558,928 & 48,681\\
PonosQ  & quiescent & 0.74 & $6.5\times 10^{12}$ & 489.6 & 265.3 & 55.0 & 9.06  & 154,489,668 & 26,521\\
\enddata
\tablecomments{Columns 3, 4, 5, 6, 7, 8, 9, and 10 give the formation redshift, present-day virial mass, virial radius, maximum circular velocity, scale radius 
and concentration parameter of the best-fit \citet{navarro+97} (NFW) spherically-averaged density profile, the number of dark matter particles 
and the number of self-bound subhalos within the virial radius of the target halos, respectively. }
\end{deluxetable*}


\subsection{Halo finding, merger-trees, and density profiles} \label{halo_detection}

Dark matter field halos and subhalos were identified with the {\sc amiga halo finder} \citep[AHF;][]{gill+04,knollmann+09}. AHF employs a recursively 
refined grid to locate local overdensities in the density field. The identified density peaks are treated as centers of prospective halos, 
subhalos, etc. An iterative procedure is then used to collect those particles that are bound to those centers while removing those that are unbound.
A bound clump of mass $m$ at a distance $d$ from the center of a more massive, $M>m$, halo is a subhalo of $M$ if $d < R + s r$, where $r$ and $R$ are the 
virial (or tidal) radii of $m$ and $M$, respectively, and $s = 0.75$ is a superposition parameter. Note that, according to this criterion, clumps 
whose centers lie just outside $R$ can still be subhalos of $M$.  
We have checked the reliability of the identification of subhalos by comparing the AHF results with those produced by Rockstar \citep{behroozi+13}.
We find nice agreement between the two halo finders in terms of subhalo mass functions, positions, and distributions of virial (tidal) radii\footnote{By default, Rockstar does not output the tidal radius of a halo. We extended the code by computing the tidal radius as the distance from the halo center of the farthest bound particle.}.

\begin{figure}
\epsscale{1.1}
\plotone{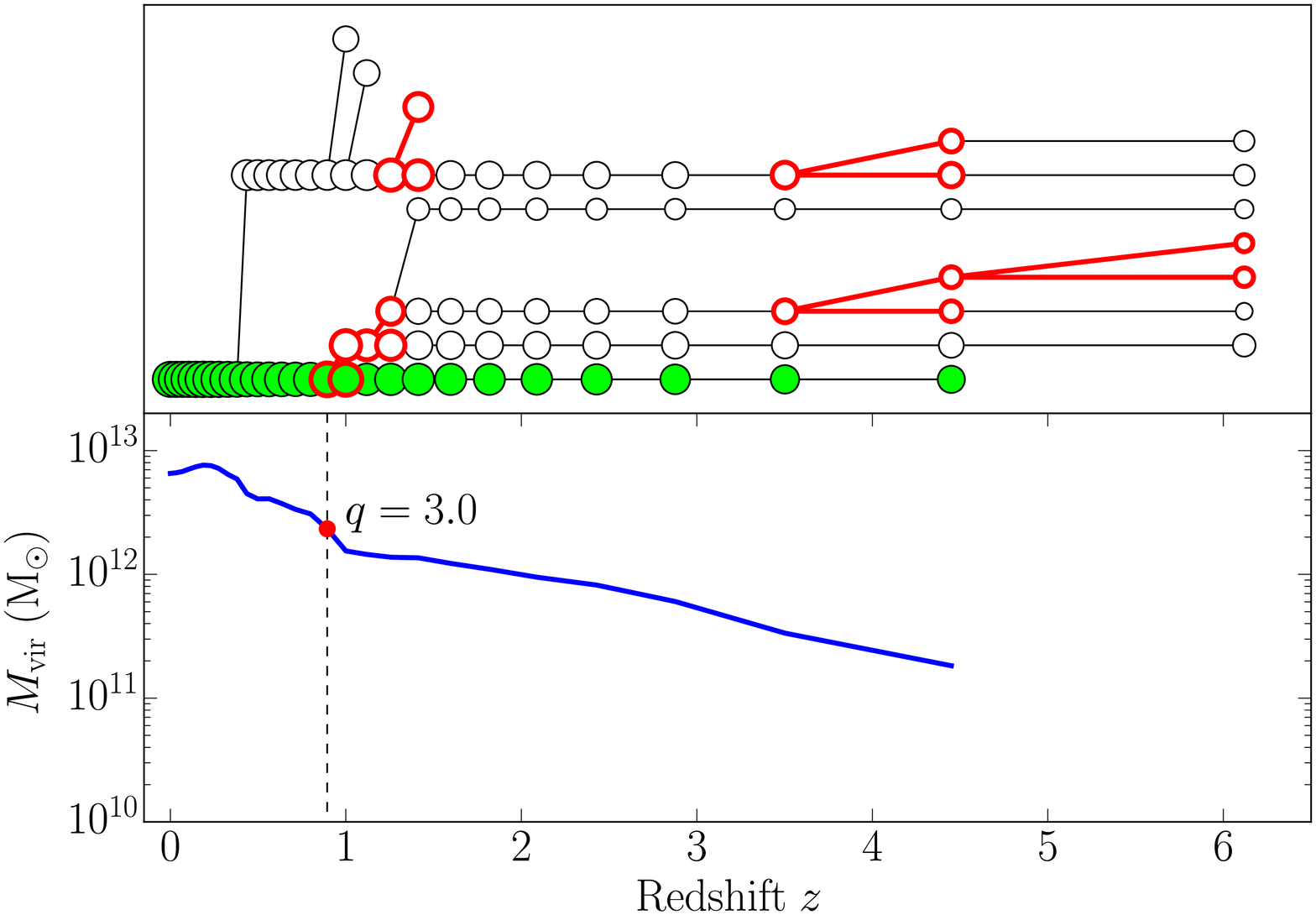}
\plotone{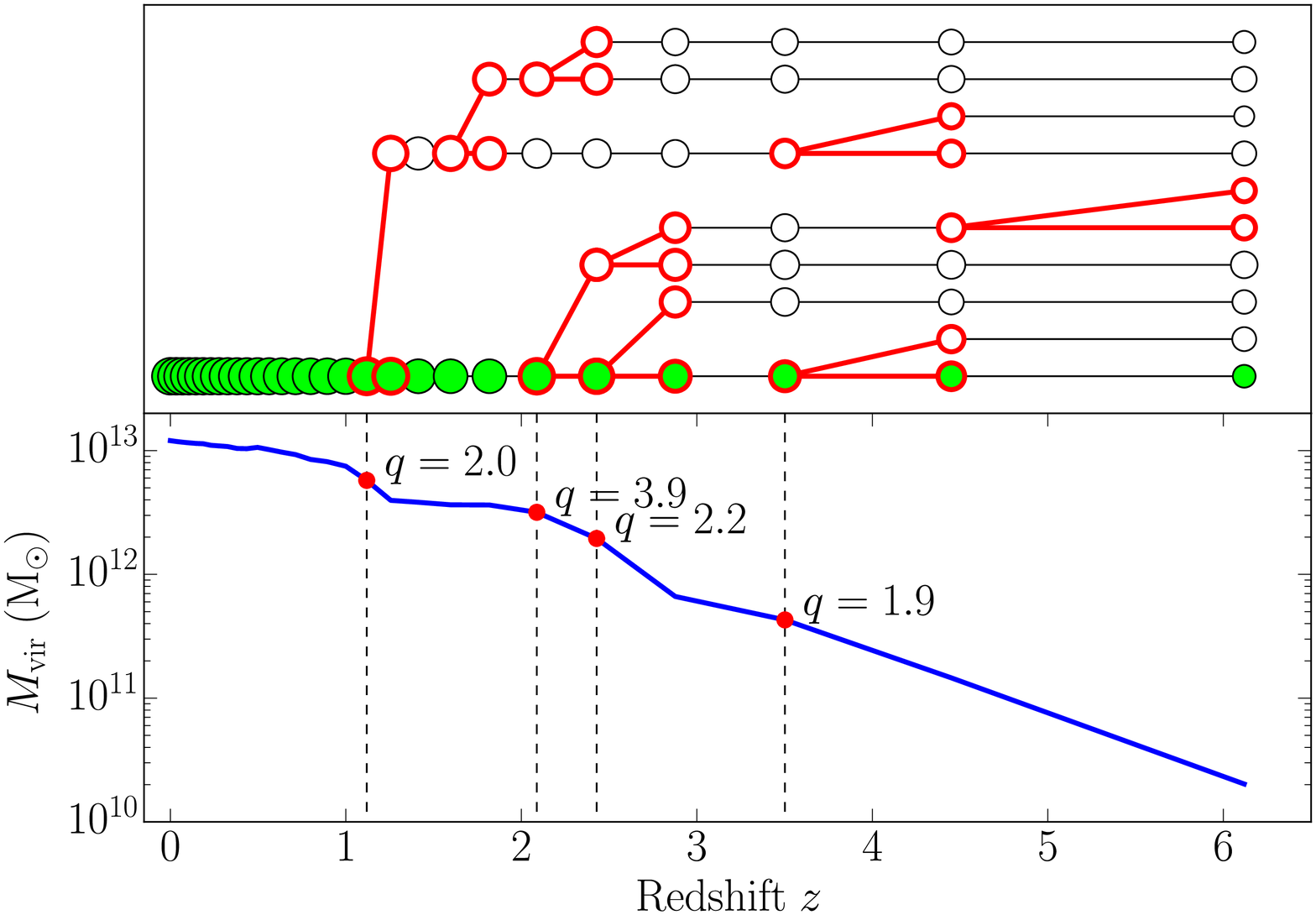}
\caption{The ``merger trees" of our simulated elliptical-sized halos, PonosQ (top) and PonosV (bottom). The main halo at the present day is plotted in the bottom-left corner, 
and all its progenitors (and also their histories) are depicted backward in time.  The size of the symbols scales with halo mass, and the ``main branch" is 
colored in green. Only mergers with mass ratios $q\equiv M/m<8$ are shown, and all major mergers with mass ratios $q<4$ are highlighted with the red color.       
The bottom portion of each panel shows the mass growth history of the main host, with major mergers marked by the vertical dashed lines.\label{fig1}}
\end{figure}

Figure \ref{fig1} shows the merger trees of the two main hosts, including only mergers with mass ratios $q\equiv M/m<8$.  We traced subhalos backward and forward in time, 
matching particles with the same ID inside each system for all snapshots between redshift 10 and 0. The main progenitor of a parent halo in a given snapshot was identified as 
the halo in the previous snapshot that maximizes the ratio $N_s/\sqrt{N n}$, where $N$ and $n$ are the number of particles in the parent and progenitor, 
respectively, and $N_s$ is the number of particles they share \citep[e.g.,][]{fiacconi+15}. As already mentioned above, PonosV and PonosQ have rather different assembly histories. 
While more massive, PonosV forms earlier (its formation redshift, defined as the redshift at which the halo had already assembled half of its mass, is $z_f=1.1$)
and undergoes four major mergers with $q<4$ at $z\lesssim  3.5$. PonosQ forms a bit later ($z_f=0.74$) and has only one major merger (with $q=3$) at $z\simeq 0.9$.
After $z=3$, PonosV mostly grows within a main dark matter filament that routes the mass accretion toward the halo.
On the other hand, PonosQ assembles around $z\sim 1-2$ after the merger of a net of short dark matter filaments converging to the position of the halo.
The latter grows quickly after $z \sim 1$ and gathers most of the nearby smaller halos, cleaning out the surroundings consistently with the more stringent 
isolation criterion adopted for the selection.

\begin{figure}
\epsscale{1.1}
\plotone{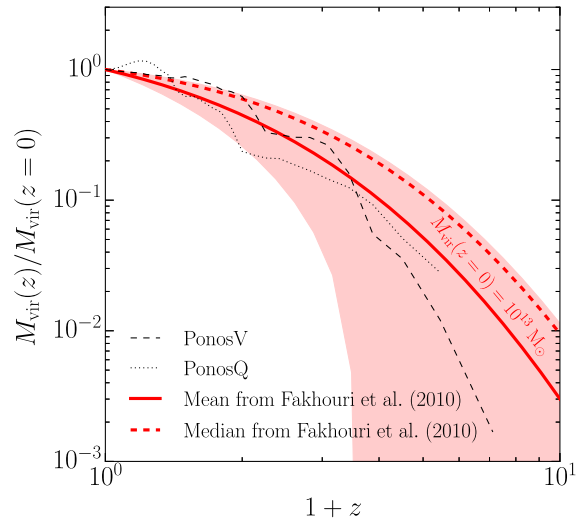}
\caption{Mass growth history of both PonosQ (thin dotted line) and PonosV (thin dashed line) as a function of redshift $z$.
Red thick lines show the mean (solid) and median (dashed) mass growth of a $10^{13}$~M$_{\odot}$ halo at $z=0$ as determined by \citet{fakhouri+10}.
The red shaded region shows the estimate of the standard deviation on the relation.
The comparison reveals that both PonosQ and PonosV have a fairly representative (though different) assembly history, as they are both compatible with the mean growth of a $10^{13}$~M$_{\odot}$ halo.
\label{fig2}}
\end{figure}

Figure \ref{fig2} compares the mass growth of PonosQ and PonosV with the mean and median growth of a $10^{13}$~M$_{\odot}$ halo at $z=0$.
These quantities have been calculated from the $z$-dependent mass accretion rates as inferred from the data of both the Millennium and Millennium-II simulation \citep[see equation (2) of][]{fakhouri+10}.
PonosV follows the average growth after $z\sim3$, while PonosQ mostly stays slightly below the curve (since it is lighter), but then it gets closer to the fit after $z\sim0.9$, as it assembles later than PonosV.
Although we do not know the dispersion on the fitted relations, we can roughly estimate the standard deviation $\sigma$ from the relation $|m - \mu|/\sigma \leq \sqrt{3/5}$, where $\mu$ and $m$ are the mean and median, respectively \citep{basu+97}.
Figure \ref{fig2} shows that both PonosQ and PonosV have mass growths compatible with the average at the present-day reference mass $10^{13}$~M$_{\odot}$; this generally qualifies both of them as ``representative'' halos for their mass scale, though they behave rather differently in detail and that influences the properties of subhalos (see Section \ref{sec3} and \ref{sec4}).

\begin{figure}
\epsscale{1.1}
\plotone{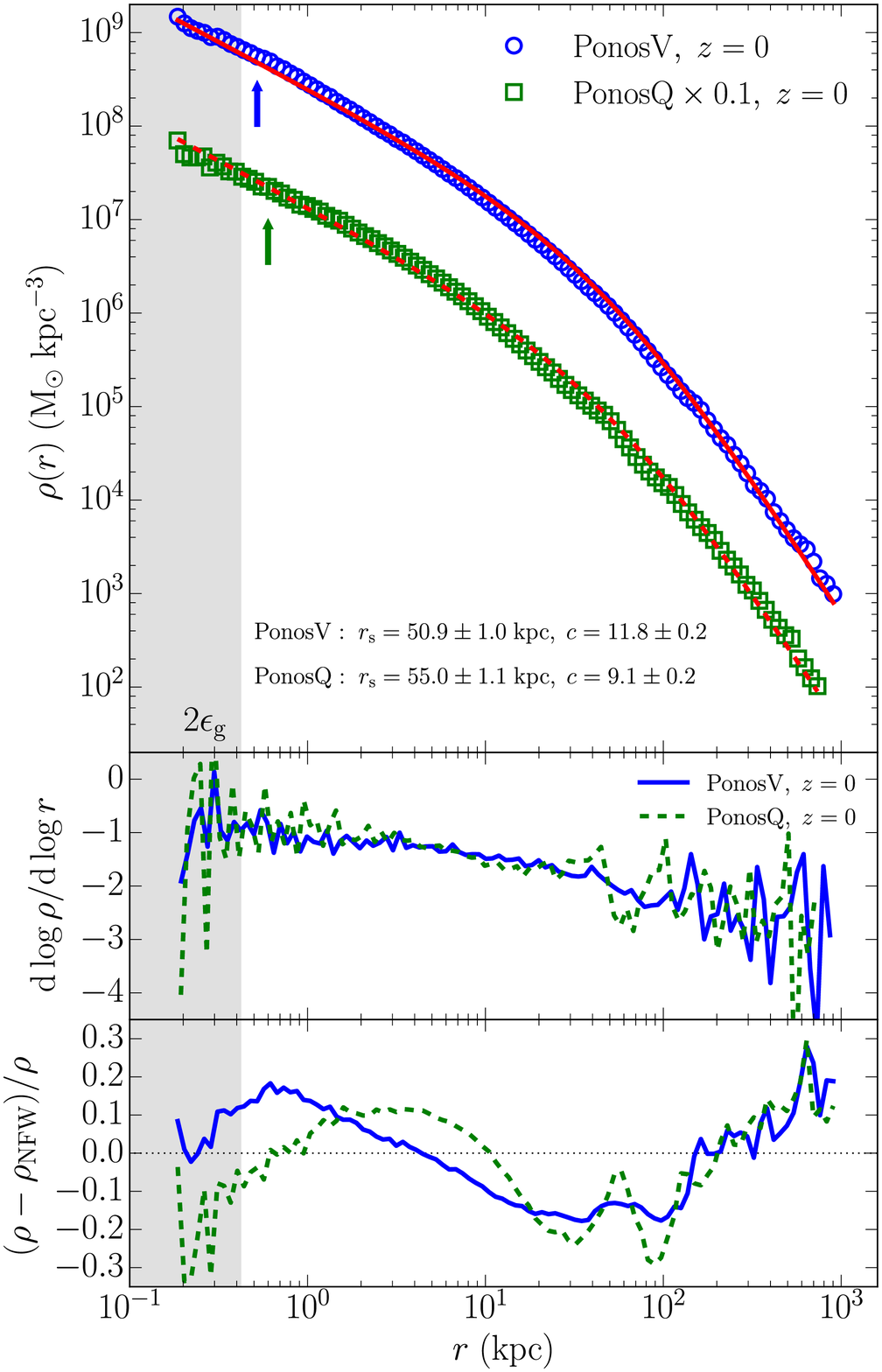}
\caption{Top: present-day density profiles of PonosV (blue circles) and PonosQ (green squares). The red solid and dashed lines are the best-fit NFW profiles. 
The density of PonosQ has been rescaled by a factor of 1/10 for clarity. The colored arrows indicates the regions within which numerical convergence is not 
achieved because of two-body relaxation \citep{power+03}. Middle: logarithmic slope of the matter density profiles of the two target halos. 
Bottom: the residuals between the density profile and the best-fit NFW profile, as a function of radius.  In all panels, the gray shading indicates the 
region inside two gravitational softening lengths.
\label{fig3}}
\end{figure}

Figure \ref{fig3} shows the present-day spherically-averaged density profiles, their logarithmic slope $d\ln\rho/d\ln r$, and the fractional deviations from the 
best-fit NFW profiles for the two target halos. The NFW formula with scale radii of 51 and 55 kpc and concentration parameters 
of 11.8 and 9.1 for PonosV and PonosQ, respectively, produces a reasonable approximation to the density profiles down to the convergence radius of each simulation. 
At all radii between 1 and 500 kpc, deviations from the best-fit NFW matter densities are typically less than 20\%.

\section{Substructure Abundance} \label{sec3}

\begin{figure}
\epsscale{1.1}
\plotone{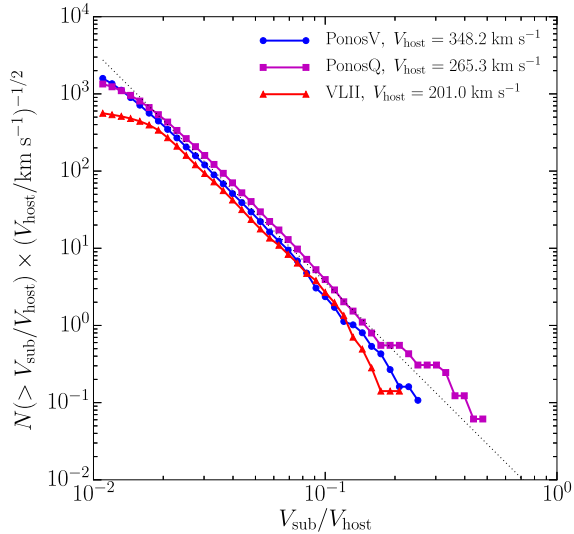}
\caption{Cumulative subhalo abundance at $z=0$ as a function of maximum subhalo circular velocity, $V_{\rm sub}\equiv \max\{\sqrt{GM_{\rm sub}(<r)/r}\}$, in units of the maximum 
circular velocity of the main host, $V_{\rm host}$. We show results for PonosV and PonosQ and, for comparison, for the Via Lactea II simulation.
The dotted line shows the best-fit power-law relation of Equation (\ref{eq:Bolshoi}). 
\label{fig4}}
\end{figure}

We counted all substructures down to a minimum of 20 bound particles, corresponding to a minimum subhalo mass of $M_{\rm res}=8.4 \times 10^{5}\,\msun$. Our selection 
results in a sample of 48,681 individual subhalos at $z=0$ in PonosV, and 26,521 subhalos in PonosQ. In Figure \ref{fig4} we present
the cumulative maximum circular velocity function (normalized to the maximum circular velocity of the host) of the subhalo population of each Ponos host. 
These are compared with the subhalo velocity functions of the Via Lactea II simulation \citep{diemand+08}.
Over the interval $0.02<x<0.08$, all velocity functions can be approximated by the power-law 
\begin{equation}
N(>x)=3.6\times 10^{-3}\,x^{-3}\,V_{\rm host}^{1/2}, 
\label{eq:Bolshoi}
\end{equation}
where $x\equiv V_{\rm sub}/V_{\rm host}$ and $V_{\rm host}$ is given in units of $\kms$ \citep[c.f.,][]{klypin+11}. Residuals from 
the power-law regression line are typically of order 20\%.  At a given $V_{\rm sub}/V_{\rm host}$, the number $N(>x)$ of subhalos 
is 56\% higher in PonosV compared to Via Lactea, and 86\% higher in PonosQ.
This is because more massive 
ellipticals are dynamically younger than Milky Way-sized galaxies, and have more subhalos that manage to survive tidal destruction. Note the relatively
large deviations above $x\gtrsim 0.15$ that are present in all three hosts compared to the fitting formula. 

\begin{figure}
\epsscale{1.1}
\plotone{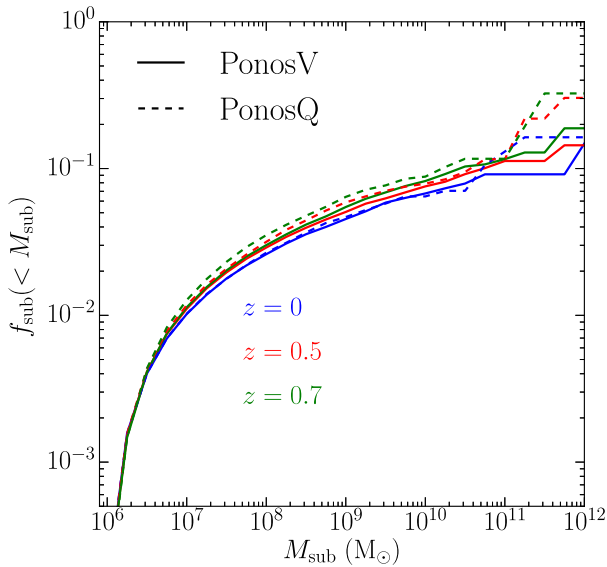}
\caption{The cumulative substructure mass fraction of PonosQ (dashed lines) and PonosV (solid lines), as a function of subhalo 
mass $M_{\rm sub}$, at $z=$0 (blue), 0.5 (red), and 0.7 (green). 
\label{fig5}}
\end{figure}

Figure \ref{fig5} shows the cumulative mass fraction in self-bound substructures of PonosQ and PonosV as a function of $M_{\rm sub}$,  
\begin{equation}
f_{\rm sub}(<M_{\rm sub})= \frac{1}{M_{\rm vir}} \int_{M_{\rm res}}^{M_{\rm sub}}m~\frac{dN}{dm}~dm, 
\end{equation}
at three different redshifts, $z=0,0.5,0.7$. Here, $dN/dm$ is the subhalo mass function.
At the present epoch, we measure a total mass fraction $f_{\rm sub}=15$\% in the violent host and $f_{\rm sub}=16.5$\% in the quiescent one. At $z=0.7$, these 
fractions increase to 19\% and 33\%, respectively, as more recently accreted subhalos are more likely to survive tidal 
stresses. Compared to PonosV, PonosQ forms later, is less concentrated, and undergoes the majority of its mergers at lower redshifts. 
As a consequence, there is less time for the orbital decay and mass loss of infalling satellites, and its substructure mass fraction is higher \citep[e.g.,][]{chen+11}. 

The definition of virial radius, while formally meaningful, is nevertheless rather arbitrary. We have checked that companion systems just 
outside the virial radius (in the spherical shell $R_{\rm vir}<r<2R_{\rm vir}$) would make  a relatively modest contribution to the ``subhalo" mass fraction, 
corresponding to less than a 30\% increase in $f_{\rm sub}$. This is not true, however, for subhalos in the shell $R_{200}<r<R_{\rm vir}$. 
To facilitate comparison with previous simulations, we have computed $f_{\rm sub}^{200}$, the subhalo
mass fraction within $R_{200}$, the radius with mean enclosed overdensity equal to 200 times the critical value\footnote{At $z=0$, we 
measure $R_{200}=444.4$ and 359.6 kpc for PonosV and PonosQ, respectively.}. At the present
epoch, we derive $f_{\rm sub}^{200}=$6.9\% in PonosV and $f_{\rm sub}^{200}=$11.4\% in PonosQ. 
These values are consistent with the mean mass fraction inferred at $z=0$ in the six Milky Way-sized halos simulated 
at resolution level 2 as part of the Aquarius project, $\langle f_{\rm sub}^{200}\rangle =(7.2\pm 2.3)$\%, as reported by 
\citet{xu+09}. However, subhalos within $R_{200}$ account for only 40\% (PonosV) and 57\% (PonosQ) of the total subhalo mass fraction $f_{\rm sub}$ within $R_{\rm vir}$,
the remaining substructure material being located between $R_{200}$ and $R_{\rm vir}$.

According to \citet{xu+09}, the scatter among the six Aquarius halos is much larger than the differences between halos at 
$z=0$ and $z=0.6$. By contrast, the two Ponos hosts have a larger mass fraction in substructure at $z=0.7$ than at present:
$f_{\rm sub}^{200}=$12.1\% in PonosV and $f_{\rm sub}^{200}=$21.2\% in PonosQ. These values are 
twice as large as measured at $z=0$. 
These large variations are associated with the late accretion of a few, relatively massive, $M_{\rm sub}\gtrsim 10^{11}\,\msun$ 
satellites. At $z=0.7$ there are two such systems in the outer halo of PonosV, and four in PonosQ.
By $z=0$, only one of them survives tidal stripping and remains above $10^{11}\,\msun$ in PonosV, and two in PonosQ. 
We note, in passing, that late minor mergers with progenitors mass ratios $q<30$ are fairly typical for descendants of 
mass $M_{\rm vir}\approx 10^{13}\,\msun$ today: an analysis of the joint data set from the Millennium and Millennium-II simulations
gives rates for such events that exceeds 1 Gyr$^{-1}$ per halo at redshift 1 \citep{fakhouri+10}.

Over the interval $10^8\,\msun<M_{\rm sub}<10^{11}\,\msun$, our measured subhalo mass function is well approximated by the power law 
\begin{equation}
\frac{dN}{d\ln M_{\rm sub}}=N_0 \left( \frac{M_{\rm sub}}{m_0}\right)^{-n}. 
\end{equation}
At $z=0$, we derive for PonosV a best-fit slope of $n=0.877$ and an amplitude at the pivot mass, $m_0=10^8\,\msun$, of $N_0=847$. 
At $z=0.7$, the best-fit parameters are $n=0.915$ and $N_0=868$. A similar relation holds also for PonosQ, with slopes and amplitudes $(n,N_0)=(0.754,362)$ 
at $z=0$, and $(n,N_0)=(0.949,344)$ at $z=0.7$. In the redshift interval $0\le z\le 0.7$, the most massive subhalo reaches 8\% of the mass of the host. 
The expected total mass fraction in self-bound substructure below our nominal resolution limit of $M_{\rm res}=8.4\times 10^5\,\msun$ is 
\begin{equation}
f_{\rm sub}(<M_{\rm res})=\frac{N_0 m_0}{M_{\rm vir}(1-n)}\left( \frac{M_{\rm res}}{m_0} \right)^{1-n}
\end{equation}
(assuming a thermal free-streaming mass limit $\rightarrow 0$). The above best-fit power laws yield between one and three percent 
of the virial mass in unresolved subhalos at $z=0$, considerably smaller compared to the fractional mass in substructures that are already resolved in our simulations.
The substructure mass fraction converges much more slowly at $z=0.7$, however, as the slope of the subhalo mass function is closer to $n=1$.   
Resolved subhalos do not trace the matter distribution of the host: tidal disruptions are most effective in the 
inner halo, leading to an antibias in the abundance profile of substructures relative to the smooth background \citep{diemand+07}. 
At the present epoch, the subhalo mass fraction within the inner 10 kpc drops to $2\times 10^{-4}$ and $1.3\times 10^{-6}$ in the violent 
and quiescent host, respectively. We caution, however, about resolution effects (numerical ``overmerging") on this quantity.
Projecting the subhalos relative to the total mass along the line-of-sight (see next section) will be less strongly influenced by the incompleteness of the inner regions of the host.


\section{Substructure Surface Mass Density} \label{sec4}

\begin{figure*}
\epsscale{1.1}
\plotone{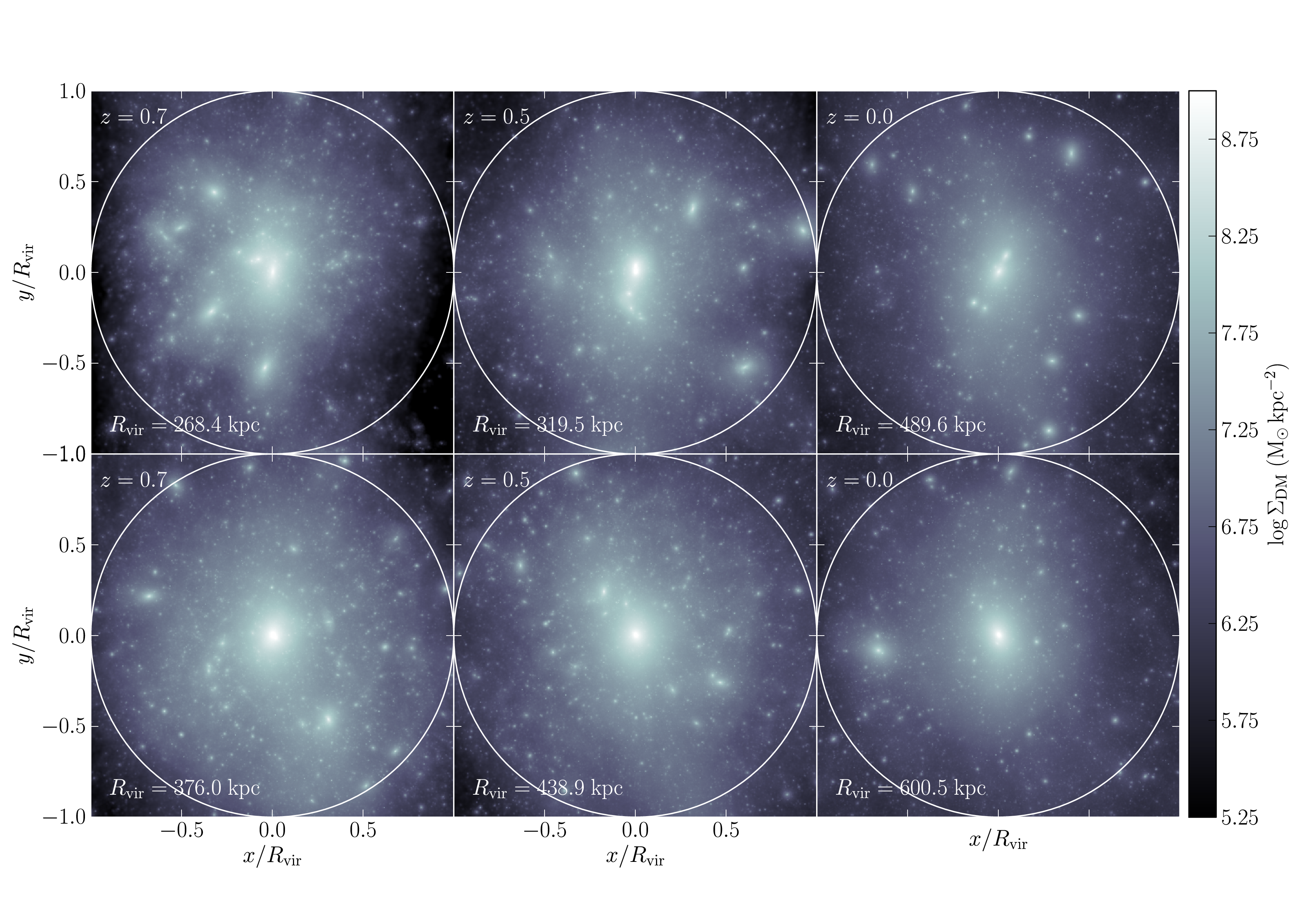}
\caption{Projected dark matter density at $z=0,0.5,0.7$ (from right to left) of PonosQ (top) and PonosV (bottom) in a slice of thickness $2R_{\rm vir}$ 
centered on the target halos. The image brightness is proportional to the logarithm of the total dark matter surface density along the line-of-sight, $\Sigma_{\rm DM}$.
The circles mark the virial radius $R_{\rm vir}$.
\label{fig6}}
\end{figure*}

Strong gravitational lensing occurs when the surface mass density along a given sightline exceeds a certain critical value, and is associated 
with high magnifications, multiple images, Einstein arcs and rings in the lens plane. 
Figure \ref{fig6} shows the surface dark matter mass density at $z=0,0.5,0.7$ of PonosQ and PonosV in a slice of thickness equal to $2R_{\rm vir}$
centered on the target halos. As expected in $\Lambda$CDM, the two Ponos halos are teeming with self-bound subhalos
on all resolved mass scales. To shed light on whether CDM substructures can account for the observed lensing flux ratio anomalies,
we have measured in the simulations the surface mass fraction in subhalos within 20 different azimuthal annuli equally spaced in $\log (R/R_{\rm vir})$ 
from $-2$ to 0, 
\begin{equation}
F_{\rm sub}(R)=\frac{\Sigma_{\rm sub}(R)}{\Sigma_{\rm DM}(R)},
\label{eq:Fsub}
\end{equation}
where $R$ is the projected radius measured from the host center, and $\Sigma_{\rm sub}$ and $\Sigma_{\rm DM}$ are the substructure and total 
dark matter projected mass densities, respectively. The surface density in substructures is computed by summing up all the subhalo bound mass 
that falls within the relevant annulus. To be specific, we adopt the following procedure to compute $F_{\rm sub}$.
First, we determine a line-of-sight by sampling uniformly the solid angle $4\pi$. This is achieved by randomly
drawing the azimuthal angle $\phi$ from a uniform distribution in the interval $[0,2\pi)$, and
the co-latitudinal angle $\theta$ from the distribution $\mathcal{P}(\theta) = \sin(\theta)/2$.
Then, we determine the line-of-sight through the versor $\mathbf{n} = (\sin\theta~\cos\phi,
\sin\theta~\sin\phi, \cos\theta)$. We calculate the projected distance of a particle in the plane
perpendicular to $\mathbf{n}$ as $R = \sqrt{|\mathbf{r}|^2 - (\mathbf{r} \cdot \mathbf{n})^2}$,
where $\mathbf{r}$ is the position vector of the particle from the centre of the halo, and we
finally bin the positions of each particle, adding its mass to the proper radial bin.
We repeat this for both the subhalos only and the host halo, and we finally calculate the ratio
between the two to derive $F_{\rm sub}(R)$.

\begin{figure}
\epsscale{1.1}
\plotone{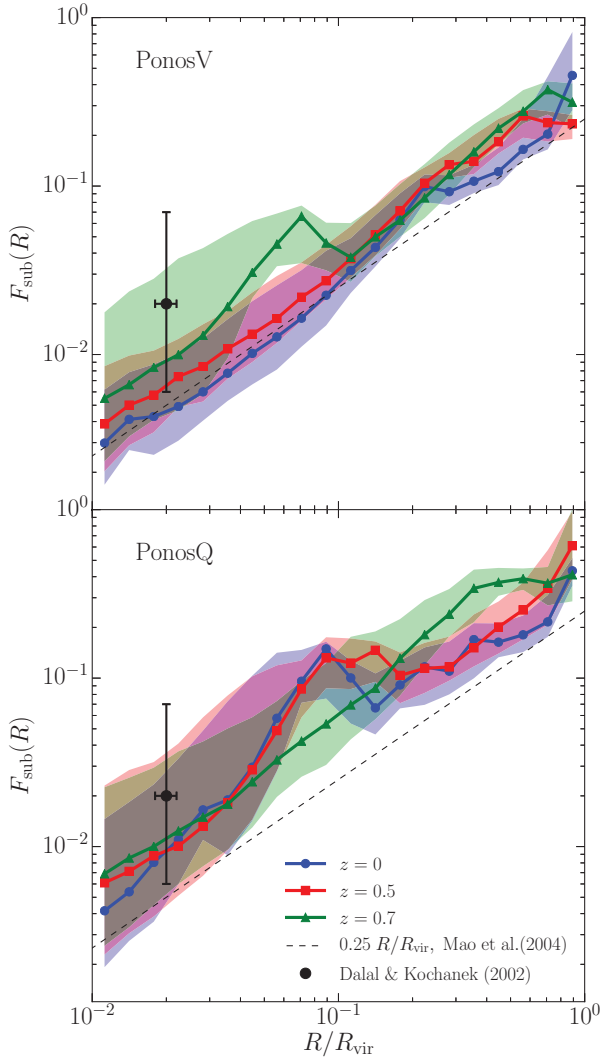}
\caption{Predicted substructure surface mass fraction at $z=0,0.5,0.7$ in PonosV (top panel) and PonosQ (bottom panel).
The panels show $F_{\rm sub}$ in azimuthal annuli as a function of the 
projected radius, $R$ (in units of the virial radius), for all subhalos belonging to the corresponding host.    
The solid lines depict the median over 300 random projections for each of the two Ponos halos, while the colored bands mark the 68\% scatter
among the different projections. The black point indicates the median and 90\% confidence level of the mass fraction in local 
substructure required to explain the flux ratio anomalies (for an assumed Einstein radius of $0.02\,R_{\rm vir}\sim $5--10 kpc).
The dashed black curve shows the results derived by \citet{mao+04} at $z=0$ from low resolution simulations of twelve halos of galactic, group, and cluster masses.
\label{fig7}}
\end{figure}

Figure \ref{fig7} shows the {\it median} subhalo surface mass fraction at $z=0,0.5,0.7$ as a function of $R/R_{\rm vir}$ over 
300 random projections for each of the two Ponos halos. The 68\% scatter among the different projections 
is marked by the colored bands, and is larger at smaller radii as the area of the azimuthal annulus around $R$ becomes progressively smaller in the inner regions. 
The value of $F_{\rm sub}$ is clearly sensitive to the assembly history of the host galaxy.
The black dashed line depicts the results derived by \citet{mao+04} at $z=0$ from simulations of twelve halos of galactic, group, and cluster masses.
Their surface mass fraction in substructures is somewhat lower than found here, perhaps because of resolution 
limitations -- the number of particles per halo in their numerical investigations is about two orders of magnitudes smaller than here.

The statistical study by \citet{dalal+02} of the anomalous flux ratios observed in a sample of seven lensed radio-loud
quasars requires $F_{\rm sub}=$0.6 to 7\% (90\% confidence, with a  median of 2\%) of the mass at the Einstein radius 
to be in substructures in order to reproduce the data.  The black point in the figure indicates the \citet{dalal+02} 
constraint -- for an assumed Einstein radius of $0.02\,R_{\rm vir}\sim$ 5--10 kpc, which appears entirely consistent
with the expectations for $\Lambda$CDM. At the Einstein radius, the {\it mean} surface mass fraction in substructure for the two 
Ponos halos ranges from 0.6\% to 2.3\%. In contrast with PonosV, there is hardly any redshift dependence in the value of this quantity for PonosQ.

These projected mass fractions are considerably higher than the mean value of 0.2\% ($0.01\le F_{\rm sub}\le 0.7\%$) 
measured at the Einstein radius in the Aquarius simulations of Milky Way-sized halos \citep{xu+09}. 
Recognizing that the substructure abundance in group-sized  halo may be larger than in 
less massive Milky Way-sized hosts, \citet{xu+15} recently 
rescaled the subhalo populations of the Aquarius suite of galactic halos and the Phoenix suite of cluster halos
to those expected in massive ellipticals. They estimate a mean surface mass fraction of substructure at the Einstein radius that is three times
bigger than in Milky Way-sized hosts, in better agreement with our findings. 
We note, however, that PonosQ has a mean subhalo surface mass fraction that is still a factor of $\sim 2$ above
the determinations by \citet{xu+15}.

At $z=0$, we find a mean substructure surface mass density around the Einstein radius of 
$\Sigma_{\rm sub}=3.6\times 10^6\,\msunkpc2$ in PonosV, and $8.9\times 10^6\,\msunkpc2$ in PonosQ. 
At $z=0.7$, we measure $\Sigma_{\rm sub}=1.4\times 10^7\,\msunkpc2$ and $1.0\times 10^7\,\msunkpc2$ in PonosV and PonosQ, respectively. 
Let us denote with $\eta$ the mean column number density of subhalos at projected 
halocentric distance $R=0.02\,R_{\rm vir}$. Over the interval $10^8\,\msun<M_{\rm sub}<10^{11}\,\msun$, 
the mean projected mass function can be written as
\begin{equation}
\frac{d\eta}{d\ln M_{\rm sub}}={\cal N}_0\left(\frac{M_{\rm sub}}{m_0}\right)^{-\ell}.
\label{eq:eta}
\end{equation}
At $z=0.7$, we measure in PonosV a best-fit slope of $\ell=0.85$ and a column number density at the pivot mass, $m_0=10^8\,\msun$, of
${\cal N}_0=0.006$ kpc$^{-2}$. In PonosQ we find a similar normalization but a steeper slope, $\ell=1.09$.
At $z=0$, we measure $\ell=0.98$ in PonosV and $\ell=0.84$ in PonosQ. Most subhalos at these image positions only appear along the line-of-sight because of projection effects. 

\begin{figure*}
\epsscale{1.1}
\plotone{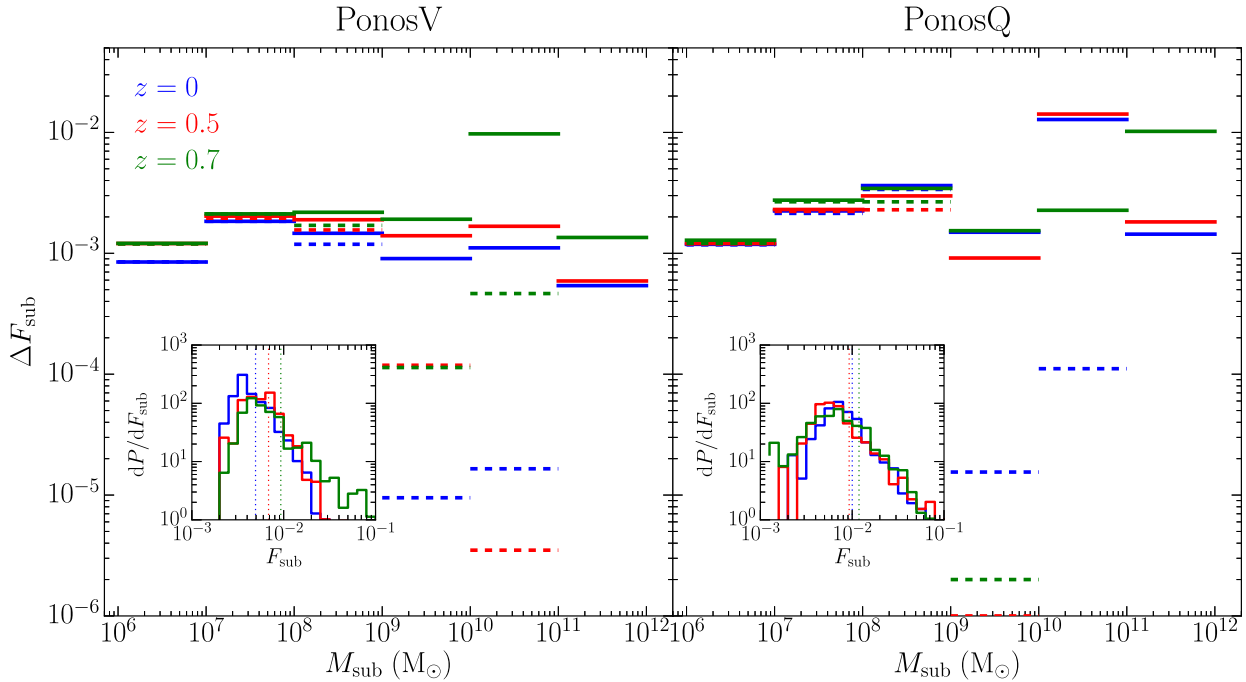}
\caption{Projected substructure mass fraction at $R=0.02\,R_{\rm vir}$ in each mass decade, for PonosV (left panel) and PonosQ (right panel).  
The individual horizontal bars depict the mean (solid lines) and median (dashed lines) over 300 random projections for each host. 
The three different redshifts inspected are listed in the left) panel. The insets show the broad probability density function 
of the total $F_{\rm sub}$, for the 300 lines of sight, at the three redshifts. Median values are indicated by the vertical dotted lines.
\label{fig8}}
\end{figure*}

In Figure \ref{fig8} we plot the {\it mean} and {\it median} surface mass fraction per decade of subhalo mass, $\Delta F_{\rm sub}$, 
again at projected halocentric distance $R=0.02\,R_{\rm vir}$. As expected from a subhalo projected mass function $d\eta/d\ln M_{\rm sub}\propto M_{\rm sub}^{-\ell}$ 
with $\ell$ close to 1, this distribution is relatively flat, with roughly equal contributions per decade of mass above the ``completeness" mass scale 
of $M_{\rm sub}=10^7\,\msun$. The contribution of subhalos with $M_{\rm sub} < 10^9\,\msun$ to the projected substructure mass fraction is between 
one fifth and one third of the total,   
with the smallest share found in the quiescent host. Massive subhalos with $M_{\rm sub}\gtrsim 10^{10}\,\msun$ only survive in the outer, $\langle r\rangle 
\gtrsim 160\,$kpc, regions of their hosts because of tidal destruction. Their presence in the projected central $\sim 10$ kpc is then typically 
associated with chance alignment. This explains the large disparity between the mean and median $\Delta F_{\rm sub}$ in the largest mass bins.
The insets in the figure  show the broad probability density function (PDF) of the total $F_{\rm sub}$ around $0.02 R_{\rm vir}$, for the 300 lines of sight. 
At $z=0.7$, the median $F_{\rm sub}$ is about 1\%, and the probability of observing values of $F_{\rm sub}>0.05$ can be as large as 12\% (PonosQ).

\begin{figure*}
\epsscale{1.1}
\plotone{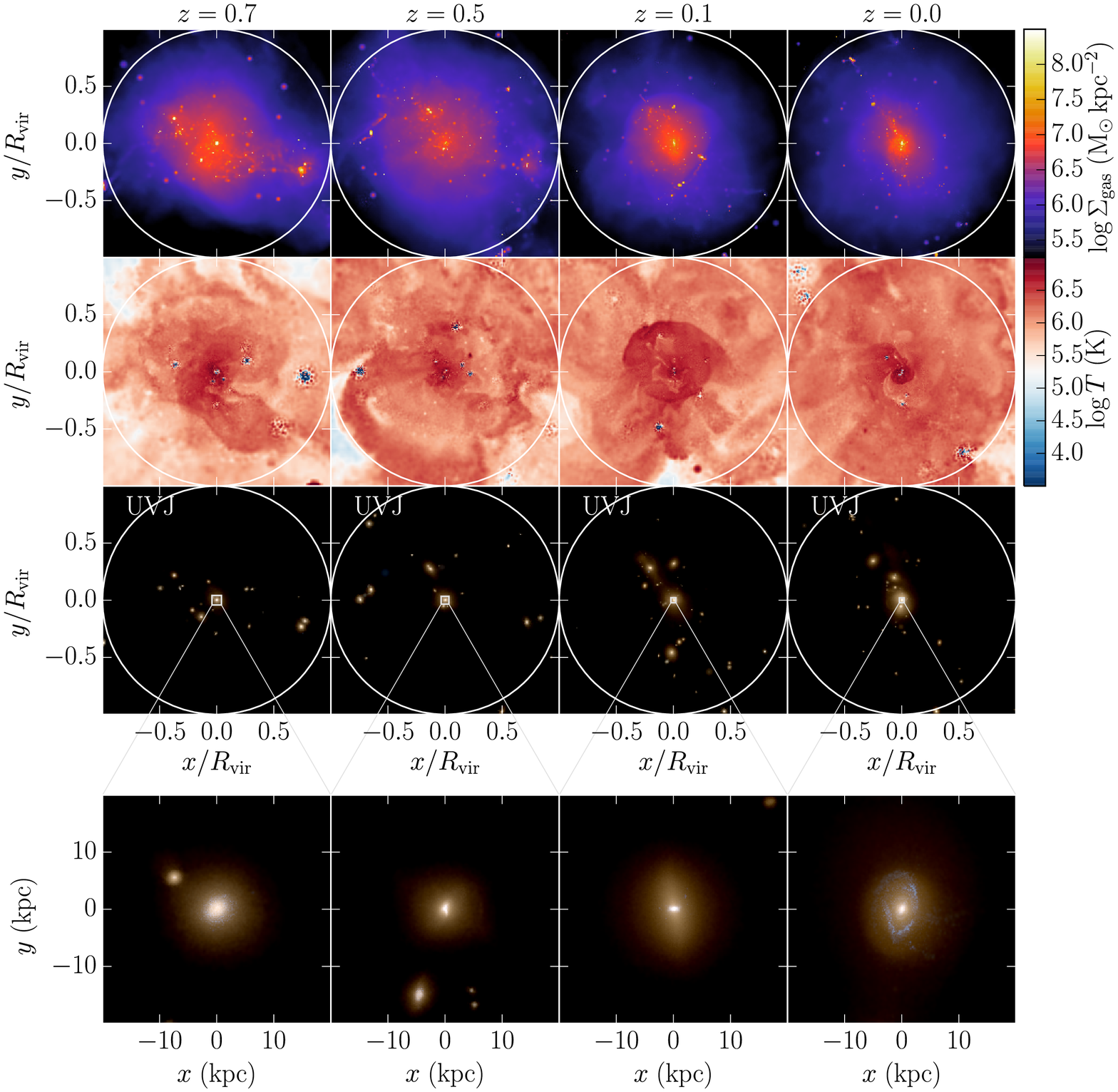}
\caption{The gas and stellar properties of PonosQH at different redshifts. From top to bottom: gas surface-density, gas temperature in a slice through the center of the host, 
mock images in the $U$, $V$, and $J$ filter bands. The top three panels encompass a physical scale equal to the virial radius, while the bottom panels shows images of the inner 40 kpc (physical).  
\label{fig9}}
\end{figure*}


\section{Baryonic Contraction}

The most severe limitation of our study is that the $N$-body very high resolution simulations used here include only dark matter. 
Standard dark matter halos are poor lenses because their central cusps ($\rho\propto r^{-1}$) are too shallow. 
In the adopted cosmology, and for a typical lens geometry with $z_{\rm lens}=0.7$ and $z_{\rm source}=2$, the critical surface density for multiple 
imaging is $\Sigma_{\rm crit}=2.2\times 10^9\,\msunkpc2$. At the same redshift and $R=0.02\,R_{\rm vir}$, our Ponos lens halos have $\Sigma/\Sigma_{\rm cr}=0.2-0.4$, 
i.e. are subcritical and unable to produce multiple images despite having maximum circular velocities close to $240\,\kms$ (PonosQ) and $340\,\kms$ (PonosV).  
Baryon cooling and condensation is expected to strongly affect the inner density profiles of halos, although predictions for these effects are far 
less certain than forecasting the distribution of dark matter in the purely dissipationless regime. As the baryons cool, they drag some of the dark matter 
inward and may even dominate the mass within one Einstein radius, thereby converting a sub-critical dark matter halo into one capable of producing multiple images.

Rather than putting in baryons ``by hand", for example by implanting an idealized model galaxy into a dark matter halo extracted from a collisionless simulation \citep{amara+06} 
or by modeling the main lens halo as a singular isothermal ellipsoid \citep[e.g.,][]{metcalf+12,xu+15}, we have followed here a different approach and run a lower 
resolution version of PonosQ (``PonosQH") with hydrodynamics using the TreeSPH code {\sc Gasoline} \citep{wadsley+04}. The code employs a 
subgrid model for the turbulent mixing of metals and energy following \citet{shen+10}.
The ICs for PonosQH  were initialized as for the collisionless run. 
The high-resolution region contains 6 million dark matter particles and an equal number of gas 
particles, for a dark matter and (initial) gas particle mass of $m_{\rm DM}=2.3\times 10^6\,\msun$ and $m_{\rm SPH}=4.5\times 10^5\,\msun$, respectively. 
This mass resolution is very similar to that of the ``small group mass" halo in the FIRE simulation suite \citep{hopkins+14}.  
The gravitational softening length was fixed to 785 pc (physical) for the dark matter, and to 501 pc for the gas.
In high-density regions the gas smoothing length was allowed to shrink to 10\% of the softening to ensure that hydrodynamic forces are resolved on 100 pc scales.
A non-thermal pressure floor \citep{agertz+09,roskar+15} was applied to stabilize scales of order the gravitational softening against gravitational collapse and avoid 
artificial fragmentation.

The simulation includes a non-equilibrium primordial chemistry network for atomic H and He, Compton cooling off the cosmic microwave background, and 
pre-computed tabulated metal-line cooling rates from the photoionization code \textsc{Cloudy} \citep{ferland+98}. A spatially-uniform, redshift-dependent cosmic 
UV background \citep{haardt+12} modifies the ionization and excitation state of the gas, photoionizing away abundant metal ions and reducing the cooling efficiency.  
Star formation proceeds at a rate ${\rm d}\rho_{\star}/{\rm d}t=0.05 (\rho_{\rm gas}/t_{\rm dyn})\propto \rho_{\rm gas}^{3/2}$ (i.e., locally enforcing the Schmidt law), where 
$\rho_{\star}$ and $\rho_{\rm gas}$ are the stellar and gas densities, and $t_{\rm dyn}$ is the local dynamical time. Star particles form in cold gas (i.e. temperature below $10^4$~K) 
that reaches a density threshold of 20 atoms cm$^{-3}$, and are created stochastically with an initial mass $m_{\star} = 1.35 \times 10^5\,\msun$ distributed following a \citet{kroupa+01} 
initial mass function.
They inject energy, mass, and metals back into the interstellar medium (ISM) through Type Ia and Type II supernovae (SNe) and stellar winds, 
following the prescriptions of \citet{stinson+06}.
A ``delayed radiative cooling" scheme for Type II SN feedback was adopted, a simplified algorithm designed to extend the Sedov-Taylor phase 
of SN remnants and mimic the effect of energy deposited in the local 
ISM by multiple, clustered sources of mechanical luminosity \citep{stinson+06}. While this approach has been found to be key in reproducing the properties 
of dwarfs \citep{governato+10,madau+14,shen+14}, late-type spirals \citep{guedes+11}, and the circumgalactic medium of $z\sim 3$ galaxies \citep{shen+13},  
many authors have recently stressed the importance of correctly accounting for the entire momentum budget of stellar feedback, including momentum injection by radiation 
pressure, stellar winds, and clustered SN explosions \citep[see, e.g.][]{agertz+13,hopkins+14,keller+14,kim+15}. 
It has been shown by \citet{agertz+13} that simulations with maximal momentum injection suppress star formation to a similar degree than found in runs 
that, like ours, adopt adiabatic thermal feedback. We note, however, that our stellar feedback model is not strictly speaking adiabatic (since SN-heated gas particles 
exchange thermal energy with the ambient medium via turbulent mixing and therefore can cool ``indirectly"), and that we did not include any feedback 
from a central active galactic nucleus (AGN). 

\begin{figure}
\epsscale{1.1}
\plotone{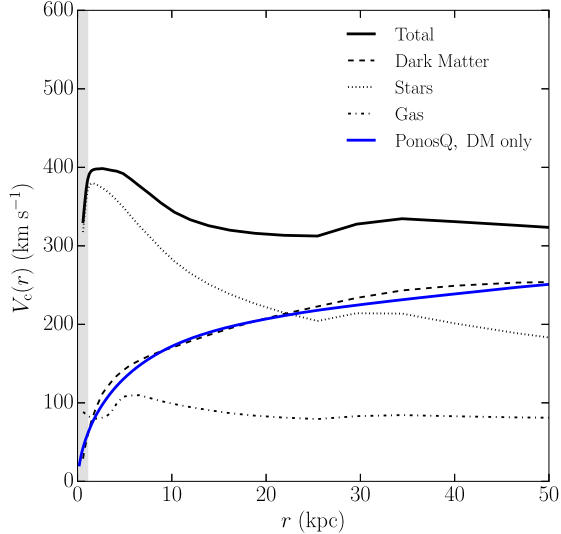}
\caption{The $z=0$ rotation curve of PonosQH, for all components (black solid line), stars (dotted line), gas (dash-dotted line), and dark matter (dashed line).
For comparison, the dark matter rotation curve (blue solid line) obtained in the purely collisionless simulation of the same object is also shown.
The gray shading indicates the region inside two gravitational softening lengths of the baryonic component.
The bump at $\sim 35$~kpc from the center is caused by a massive satellite system.
\label{fig10}}
\end{figure}

Figure \ref{fig9} shows gas surface-density and temperature maps at four different redshifts, together with mock images in the $U$, $V$, and $J$ filter bands. 
The top three panels encompass a physical scale equal to the virial radius, while the bottom panels are images of the inner 40 kpc (physical).  
Each star particle was assigned a luminosity from tables of mass-to-light ratios based on the isochrones and synthetic stellar populations of \citet{bressan+12}. 
At the present epoch, PonosQH has an early-type morphology, with $U-V=1.5$ and $V-J=0.9$ colors that are typical of red sequence galaxies.
Its present-day stellar mass within 20 kpc from the center, $M_{\star}=2.3\times 10^{11}\,\msun$, implies a star formation efficiency, $M_{\star}/M_{\rm vir}=0.035$, which is 
slightly higher than the value inferred for massive halos by \citet{kravtsov+14}, but compatible within $2 \sigma$. The surface brightness profile follows a de Vaucouleurs law with 
a $V$-band effective radius $R_{e,V}=3.6$~kpc, while the half-light radius measured directly from the light profile is $R_{1/2} = 3.1$~kpc.
The baryonic content of PonosQH is close to the universal baryon fraction. 
The mass of cold ($T<10^4\,$K) gas per unit $K$-band luminosity, $M({\rm HI})/L_K=0.03$, places this system towards the high-end of the distribution observed 
in the ATLAS survey of nearby early-type galaxies \citep{serra+12}.

\begin{figure}
\epsscale{1.1}
\plotone{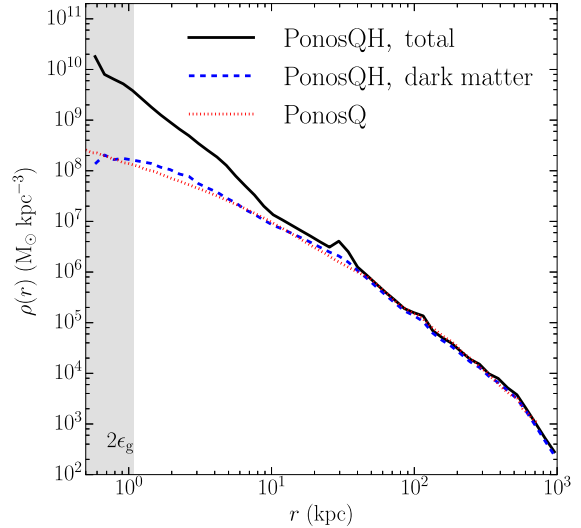}
\caption{The $z=0$ density profile in PonosQH for all components (black solid line) and dark matter (dashed line). 
For comparison, the dark matter density profile (red dotted line) obtained in the purely collisionless simulation of the same object is also shown.
The gray shading indicates the region inside two gravitational softening lengths of the baryonic component. 
We do not observe any significant evidence for adiabatic contraction or expansion in the dark matter at 
the center of the halo in the hydrodynamical run.
\label{fig11}}
\end{figure}

We now focus on the the impact of baryonic infall on the 
inner density profile and projected substructure mass fraction.
In Figure \ref{fig10} we show the $z=0$ rotation curve, $V_c(r)=\sqrt{GM(<r)/r}$, for 
all components (gas, stars, and dark matter) separately. The dark matter circular velocity curve obtained in the collisionless simulation of the 
same object is plotted for comparison. Baryons dominate within the inner 20 kpc, and the ensuing rotation curve remains relatively flat from 
several kpc to several tens of kpc.
The bump at $\sim 35$~kpc is due to the presence of a massive, $\sim 1.5 \times 10^{11}\msun$, satellite. 
The total mass within 10 kpc has increased by a factor of three relative to the purely collisionless simulation.
The effective power-law slope of the total (luminous plus dark; $\rho\propto r^{-\gamma}$) mass distribution in the range 1.5-40 kpc 
is $\gamma=2.19\pm 0.07$, i.e. 
marginally steeper than isothermal ($\gamma=2$). This value appears to be consistent with the results of the Sloan Lens ACS Survey (SLACS) of 73 early-type 
galaxies with $0.08<z_{\rm lens}<0.5$ and stellar masses above $10^{11}\,\msun$: $\langle \gamma\rangle =2.078\pm 0.027$ with an intrinsic scatter of $0.16\pm 
0.02$ \citep{auger+10}. Recent modeling of the mass density profiles of early-type galaxies also yields super-isothermal central 
slopes, with $\langle \gamma\rangle =2.15\pm 0.04$ \citep{chae+14} and $\langle \gamma\rangle =2.19\pm 0.03$ \citep{cappellari+15}. 

\begin{figure}
\epsscale{1.1}
\plotone{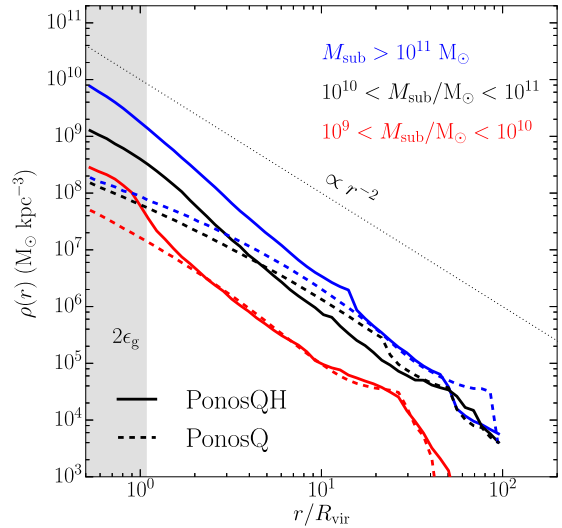}
\caption{The $z=0$ average density profiles of massive subhalos. Solid lines: PonosQH. 
Dashed lines: PonosQ. The difference between the two sets shows the impact of baryonic contraction. 
The red, black, and blue colors refer to the subhalo mass intervals $10^9<M_{\rm sub}/\msun<10^{10}$, $10^{10}<M_{\rm sub}/\msun<10^{11}$, and 
$M_{\rm sub}/\msun>10^{11}$, respectively. For reference, the dotted line shows an isothermal, $\propto r^{-2}$, profile.
The gray shading indicates the region inside two gravitational softening lengths of the baryonic component.
\label{fig12}}
\end{figure}

\begin{figure*}
\epsscale{1.1}
\plotone{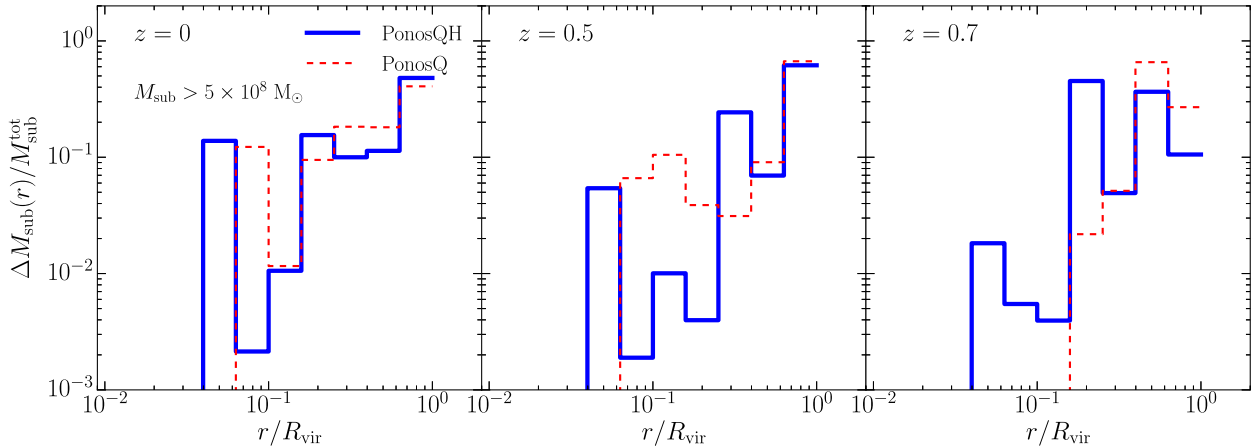}
\caption{Impact of baryonic contraction on the radial distribution of massive subhalos. Solid line: radial distribution of the cumulative mass, $\Delta M_{\rm sub}$,
in all PonosQH subhalos with  $M_{\rm sub}>5\times 10^8\,\msun$.  Dashed line: same for PonosQ. The distance to the center of the main halo is normalized to the 
virial radius. The subhalo mass is normalized to the total mass in subhalos $>5\times 10^8\,\msun$ identified in the hydro and collisionless simulations. 
The left, middle, and right panels show the distribution at three different redshifts.
\label{fig13}}
\end{figure*}

The steeper inner cusp is clearly seen in the density profile plot (Figure \ref{fig11}). In response to the slow addition of baryons to the center, 
dark matter may be pulled inwards through a process known as adiabatic contraction \citep{blumenthal+86,gnedin+04,pillepich+14}. 
Other processes may cause the dark matter to expand instead,
such as the transfer of orbital energy via dynamical friction following dry minor mergers \citep{el-zant+01} and rapid gravitational potential fluctuations tied 
to efficient SN feedback \citep{pontzen+12,madau+14}. At the present-epoch, PonosQH shows little evidence for strong dark matter contraction or expansion. Using a generalized 
NFW model, we find an inner dark matter density slope, $\gamma_{\rm DM}=1.1\pm 0.1$, which is consistent with unity.  
The mean dark matter mass fraction projected within a cylinder of radius equal to the Einstein radius ($0.02\,R_{\rm vir}$) is 
$F_{\rm DM}(<R_{\rm vir})=0.38 \pm 0.01$, in agreement with the average value, $0.38\pm 0.07$, measured in the SLACS lens sample by \citet{bolton+08}.

Baryonic contraction acts both for the host halo and for its subhalos. At $z=0$, we identify in PonosQH more than 400 subhalos, 285 of which are
above $10^8\,\msun$ (the minimum subhalo mass corresponding to 20 bound dark matter particles is now $M_{\rm res}=4.6\times 10^7\,\msun$).    
Figure \ref{fig12} shows the $z=0$ density profiles of massive, $M_{\rm sub}>10^9\,\msun$ subhalos averaged over different mass intervals. 
All subhalos above $M_{\rm sub}>10^{10}\,\msun$ form stars and are 
therefore ``luminous", while half of those in the range $10^9<M_{\rm sub}/\msun<10^{10}$ are dark.
The difference between PonosQH and the collisionless PonosQ is clear. Baryonic infall increases the total mass within, say, 2 kpc 
from the satellite center by a factor that ranges from 
2 ($10^9<M_{\rm sub}/\msun<10^{10}$) to 14 ($M_{\rm sub}/\msun>10^{11}$) relative to the purely collisionless simulation. 
As a consequence, the mean maximum circular velocity of subhalos in the mass range $10^9<M_{\rm sub}/\msun<10^{10}$ rises 
from $28\,\kms$ to $49\,\kms$, and from $83\,\kms$ to $122\,\kms$ in the case of $10^{10}<M_{\rm sub}/\msun<10^{11}$ subhalos.
Similar effects of baryonic contraction in massive satellites of Milky Way-sized halos have also been recently reported by \citet{zhu+15}. 
And while the slope of the internal density profile of subhalos may have little effect on the frequency of flux anomalies \citep{metcalf+12}, 
baryon cooling and condensation within massive subhalos will make them more resilient to tidal disruptions \citep[e.g.][]{maccio+06}. 

In Figure \ref{fig13} we compare the radial distribution of the cumulative subhalo mass in PonosQH and PonosQ. 
Only subhalos with $M_{\rm sub}>5\times 10^8\,\msun$ are included in the comparison.
The distance to the center of the main halo is normalized to the virial radius (slightly larger in PonosQH), 
while the subhalo mass is normalized to the total mass in all $M_{\rm sub}>5\times 10^8\,\msun$ subhalos identified in the hydro and collisionless simulations.
The left, middle, and right panels show subhalos at three different redshifts. The substructure mass in the inner regions is consistently larger 
in PonosQH. This notwithstanding the fact that the increased mass concentration at the center of the main host induces stronger tidal forces that can 
potentially destroy subhalos. At $z=0.7$, for example, there are 4 subhalos more massive than $10^9\,\msun$ within 30 kpc of the center of PonosQH,  
compared to none in PonosQ. At $z=0.5$, the innermost $M_{\rm sub}>10^{10}\,\msun$ satellite in PonosQH is at 17 kpc from the center, versus 46
kpc for PonosQ.  

\begin{figure}
\epsscale{1.1}
\plotone{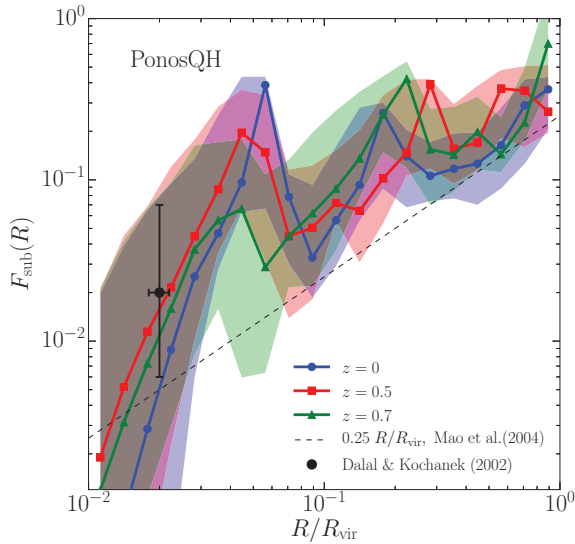}
\caption{Impact of baryonic contraction on the substructure surface mass fraction. Here, we have measured the total (baryons $+$ dark matter) substructure 
surface density in the hydro simulation PonosQH. Legend is the same as Figure \ref{fig7}. Because PonosQH only resolve substructures at intermediate and  
large mass scales, the sightline-to-sightline scatter has increased compared to Figure \ref{fig7}. 
\label{fig14}}
\end{figure}

The effect of baryon cooling and condensation on the predicted projected mass fraction is depicted in Figure \ref{fig14}. We have modified the numerator and 
denominator of Equation (\ref{eq:Fsub}) to account for the impact of baryons on the main halo and subhalos as
\begin{equation}
F_{\rm sub}(R)=\frac{\Sigma_{\rm sub}(R)}{\Sigma_{\rm DM}(R)+\Sigma_{\rm bar}(R)}.
\end{equation}
Here, $\Sigma_{\rm sub}$ is the substructure projected mass (baryons $+$ dark matter) density, while
$\Sigma_{\rm DM}$ and $\Sigma_{\rm bar}$ are the total dark matter and baryon surface densities of the main host, all measured in PonosQH.
{\it PonosQH does not have enough resolution to generate a realistic substructure population at masses $M_{\rm sub}\lesssim  5\times 10^8\,\msun$, 
and so this figure should be interpreted as the surface mass fraction in substructure at intermediate and large mass scales only.}
As in Figure \ref{fig7}, we have plotted the median substructure mass fraction rather than the mean, for better comparison with the \citet{dalal+02} constraints. 
Note how, even without small-scale power, the predicted median at the Einstein radius and $z\gtrsim 0.5$ is consistent with the data. 


\section{Discussion and Conclusions}

In this Paper, we have used new very high resolution $N$-body and hydrodynamical cosmological simulations of early-type massive galaxies 
to refine $\Lambda$CDM predictions for substructure gravitational lensing. It has long been known that, if dark matter substructure 
can survive to constitute a percent or more of the surface mass density at small impact parameters, it will cause detectable anomalies in the 
magnification ratios of multiply-imaged QSOs \citep{metcalf+01}. In our dissipationless simulations, we have found that the total mass fraction 
in self-bound subhalos increases from about 15\% at the present epoch to 20-30\% at redshift 0.7. In projection, the average 
fraction of surface mass density in substructure around the Einstein radius at $z=0.7$ exceeds 2\%. More massive hosts (as well as hosts at 
higher redshifts) are predicted to be dynamically younger and therefore more clumpy, as their subhalos are accreted more recently 
and tend to survive tidal destruction. Indeed, our two Ponos $M_{\rm vir}\sim 10^{13}\,\msun$ halos have significantly higher, by as 
much as a factor of 10, mean projected substructure mass fractions at $z=0.7$ than measured at the present-epoch in the Aquarius and Via Lactea 
simulations of Milky Way-sized systems. As a result, the frequency of flux ratio anomalies predicted by the richer substructure population 
of early-type galaxy halos increases noticeably compared to estimates based on Milky Way-sized halos \citep[e.g.,][]{amara+06,maccio+06,xu+09}. 

We have incorporated the effects of baryonic contraction on the host halo using twin, lower-resolution hydrodynamical simulations that include 
metallicity-dependent gas cooling, a star formation recipe based on a high gas density threshold, and a delayed-radiative-cooling scheme for feedback by SNe, 
but no AGN feedback. The inclusion of the baryonic component produces at the present epoch a red sequence galaxy with a super-isothermal central slope 
in the total (luminous plus dark) mass distribution, a rotation curve that is relatively flat from several kpc to several tens of kpc, a projected dark matter mass fraction
inside the Einstein radius of 40\%, and little evidence for strong dark matter contraction or expansion. Such properties of the 
matter density profile appear broadly consistent with observations of early-type galaxies \citep[e.g.,][]{bolton+08,auger+10,chae+14,cappellari+15}.
Baryonic contraction increases the number of massive subhalos in the inner regions of the main host. 

Our $\Lambda$CDM simulations, both in the purely collisionless case or after accounting for the impact of baryonic contraction, 
appear at first sight to predict enough substructure to explain the frequency of lens anomalies in currently available samples \citep{dalal+02}. 
While this is indeed promising, numerical calculations of the lensing potential, deflection angles, and magnifications are needed
to generate theoretical flux-ratio probability distributions for comparison with the observations. 
Two such studies have been performed recently. \citet{metcalf+12} created a large number of simulated lenses with finite source sizes, 
compared the predicted flux-ratio probability distributions in the presence of substructure to an observational sample of seven lenses, and found 
approximate consistency with $\Lambda$CDM $N$-body simulations. \citet{xu+15} constructed samples of lens potentials by adding 
(rescaled to those expected in massive ellipticals) subhalo populations from the galaxy-scale Aquarius and the cluster-scale Phoenix simulation 
suites, and matched the resulting flux ratio distributions to the best available sample of radio lenses. They reached the conclusion that
CDM substructures cannot account for all the observed anomalies. Such detailed investigations are beyond the scope of this work, and we defer them to a future paper.
Here, we only note that, according to Figure 9 in \citet{metcalf+12}, a subhalo surface mass density of $\Sigma_{\rm sub}=10^7\,\msunkpc2$
would imply a 15\% chance of a clear outlier in the general distribution of $\Delta\theta$ and $R_{\rm cusp}$ values predicted in the
absence of substructures, consistent with the frequency (one out of seven) actually observed\footnote{Here, $\Delta\theta$ 
and $R_{\rm cusp}$ are two parameters that characterize 4-image lenses. Sources near a cusp in the caustic produce ``cusp" configurations 
with three of the images (``triplet") lying close together on one side of the lens galaxy.
The parameter $\Delta\theta$ is the angular separation between the close triplet, and is small when the source is near one of the cusps in the caustic.
In any smooth lensing potential, the three close images satisfy an asymptotic magnification relation (the ``cusp-caustic relation")
$R_{\rm cusp}=|\mu_A+\mu_B+\mu_C|/(|\mu_A|+|\mu_B|+|\mu_C|)\rightarrow 0$,
with the total absolute magnification $|\mu_A| + |\mu_B| + |\mu_C| \rightarrow \infty$ \citep[e.g.][]{keeton+03}.}.
A substructure surface mass density of $10^7\,\msunkpc2$ is comparable to that measured in the Ponos halos. Contrary to the model of \citet{metcalf+12}, however, 
most of this projected mass density is associated with subhalos more massive than $M_{\rm sub}=10^9\,\msun$.

Which brings up the next point. 
Magnifications, which depend on the second spatial derivative of the lensing potential, are affected about 
equally by all mass scales provided the Einstein radius of the deflector is larger than the size of the background source. 
The constraints provided by the frequency of flux anomalies cannot therefore discriminate between different substructure mass scales.  
On the other hand, distinguishing CDM from keV-mass WDM -- in which the free-streaming cutoff occurs on dwarf galaxy scales -- requires measuring the subhalo mass
function well below a mass of $\sim 10^9\,\msun$ \citep[e.g.,][]{li+2015}. We find that, in the Ponos hosts, the contribution of subhalos 
with $M_{\rm sub}<10^9\,\msun$ to the total 
projected mass fraction is sub-dominant, between one fifth and one third of the total. This fact highlights the potential importance of dwarf-sized systems 
in the anomalous flux ratio problem. Small dark substructures may actually play a lesser role in causing flux anomalies than the more massive subhalos, and 
some of these massive perturbers may actually contain visible dwarf galaxies. Indeed, three out of the seven gravitational lens systems used by 
\citet{dalal+02} to statistically detect dark substructure around early-type galaxies show evidence for additional mass structure in the form 
of luminous dwarf satellites \citep[see, e.g.,][and references therein]{mckean+07}. In the case of the lens systems B2045+265 and MG 2016+112, 
the dwarf perturber constitutes about 1\% of the total projected lens mass, which seems consistent with the predictions of our Ponos simulations. 
Moreover, other components of the lensing galaxy, like an edge-on disk, may also account for some anomalies, as suggested by recent observations of 
CLASS B1555+375 \citep{hsueh+16}. It remains therefore unclear how effective a probe of the matter power spectrum on sub-galactic scales the current, 
small sample of lensed radio-loud quasars may actually be. Another complicating factor makes the comparison between $\Lambda$CDM predictions
with the observations still rather preliminary. In addition to substructures within the halo of the lensing galaxy, dwarf-sized perturbers along the 
line-of-sight to the lensed quasar may also affect the lens potential and give rise to flux ratio anomalies. And while lensing by dark matter 
clumps near the lens galaxy may be more effective, the cumulative effect of all intergalactic halos along the line-of-sight could be significant 
\citep{chen+03,wambsganss+05,metcalf05,inoue+12,inoue16}. 

\begin{figure}
\epsscale{1.1}
\plotone{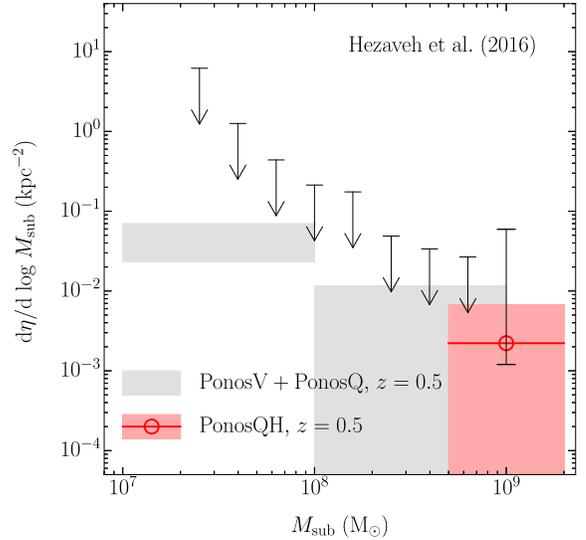}
\caption{The differential projected mass function of subhalos at the Einstein radius. 
The errorbars indicate the 95\% confidence limits on the projected number density of subhalos around 
the dusty galaxy SDP.81 \citep{hezaveh+16}. For comparison, the shaded grey rectangles show the 90\% confidence region in a 
decade of mass at $z=0.5$ from the combined PonosQ and PonosV simulation sets.
The red circle with the red error band shows the mean and 95\% confidence limits in PonosQH in the mass bin $0.5-2 \times 10^{9}\,\msun$.
\label{fig15}}
\end{figure}

With the next generation of wide-field optical surveys capable, e.g., of increasing the samples of multiply-imaged quasars by two orders of magnitude \citep{oguri+10}, 
however, strong gravitational lensing appears poised to dramatically improve our understanding of 
dark matter substructure in galaxy halos. Flux ratio anomalies are only one means of detecting substructure. The most massive 
subhalos may also visibly perturb the deflection angle of lensed images, causing ``astrometric anomalies" \citep{metcalf+01,chiba02,chen+07}. High-precision measurements
of time delay perturbations between images in strong gravitational lens systems may complement flux ratio and astrometric anomalies as they 
depend on a different moment ($M_{\rm sub}^2$) of the subhalo mass function \citep{keeton+09}. An alternative technique, the ``direct gravitational imaging" of individual 
clumps based on perturbations to the surface brightness of highly magnified Einstein rings, has been developed by \citet{koopmans05}. 
Detailed studies of individual Einstein ring systems have led \citet{vegetti+12} to detect a
$(1.9\pm 0.1)\times 10^8\,\msun$ dark subhalo in an extended optical galaxy-galaxy lens system at $z=0.88$.
From a search of mass clumps in a sample of 11 lens galaxies from the SLACS,
\citet{vegetti+14} infer a mean projected substructure mass fraction $F_{\rm sub}=0.0076^{+0.0208}_{-0.0052}$ (68\% confidence level) 
around the Einstein radius of massive early-type host galaxies at $\langle z_{\rm lens}\rangle=0.2$. 
This fraction is again consistent with the expectations from $\Lambda$CDM simulations.
Recently, \citet[][see also \citealt{inoue+16b}]{hezaveh+16} have used ALMA observations of the strongly lensed dusty galaxy SDP.81 to find evidence for a 
$M_{\rm sub}=10^{8.96\pm 0.12}\,\msun$ subhalo near one of the images, and produce constraints on the projected abundance of 
substructure near the Einstein radius. Figure \ref{fig15} shows the resulting constraints on the differential projected subhalo mass function, compared with 
predictions from the Ponos simulation sets. Again, the observations appear consistent with theoretical expectations. 
More importantly, the study by \citet{hezaveh+16} shows that future ALMA data have the potential of constraining the abundance of dark matter subhalos 
down to $M_{\rm sub}\sim 2\times 10^7\,\msun$.
Independently, \citet{woldesenbet+15} have reached similar conclusions (i.e. the necessity of $\gtrsim 10^{8}$~M$_{\odot}$ substructures projected near the Einstein radius) by using a model-free analysis of image positions in the relative angle space.

Our hydro simulations do not have enough resolution to generate a realistic substructure population at masses $M_{\rm sub}\lesssim  5\times 10^8\,\msun$, and 
higher resolution runs with dissipation are in the making in order to extend our analysis to smaller-mass subhalos. It is also necessary to perform more 
$N$-body simulations of early-type galaxy halos in order to obtain unbiased samples and reliable statistics of the subhalo population.
We plan to perform a statistical exploration of flux anomalies induced by substructures near the Einstein radius on a larger halo sample.
Since our results emphasize the relevance of dwarf-sized systems, such exploration may rely on a large suite of less challenging lower resolution simulations, as long as they can robustly resolve subhalos with masses $\gtrsim 10^{8}$~M$_{\odot}$.
This is instrumental to broadly asses the impact of substructures on the occurrence of flux anomalies in lensed systems.


\acknowledgments
We thank the anonymous Referee for useful comments that improved the quality of this work.
We thank Lucio Mayer for useful discussions.
We thank Oliver Hahn and the AGORA collaboration for help with the initial conditions of the simulations, and Yashar D. Hezaveh for providing the data plotted in Figure \ref{fig15}. 
Support for this work was provided to P.M. by the NSF through grant AST-1229745, by NASA through grant NNX12AF87G, and by the Pauli Center for Theoretical Studies. 
D.F. is supported by the Swiss National Science Foundation under grant 200021\_140645.


\bibliographystyle{apj}
\bibliography{ponos}


\end{document}